\documentclass[10pt,a4paper]{article}
\usepackage[utf8]{inputenc}
\usepackage{graphicx}
\usepackage{array}
\usepackage{hyperref}
\usepackage{amsmath}
\usepackage{caption}
\usepackage{subcaption}
\usepackage{authblk}

\usepackage{geometry}
\hypersetup{
    colorlinks=true,
    linkcolor=blue,
    filecolor=magenta,      
    urlcolor=cyan, 
}

\title{GAN-AE : An anomaly detection algorithm for New Physics search in LHC data}
\author[1]{Louis Vaslin}
\author[2]{Vincent Barra}
\author[1]{Julien Donini}
\affil[1]{LPC, Université Clermont Auvergne, CNRS/IN2P3, Clermont-Ferrand; France}
\affil[2]{Université Clermont Auvergne, CNRS, LIMOS, UMR 6158, Clermont-Ferrand; France}

\date{\today}

\begin{document}

\maketitle

\begin{abstract}
    In recent years, interest has grown in alternative strategies for the search for New Physics beyond the Standard Model.
    One envisaged solution lies in the development of anomaly detection algorithms based on unsupervised machine learning techniques.
    In this paper, we propose a new Generative Adversarial Network-based auto-encoder model that allows both anomaly detection and model-independent background modeling.
    This algorithm can be integrated with other model-independent tools in a complete heavy resonance search strategy.
    The proposed strategy has been tested on the LHC Olympics 2020 dataset with promising results.
\end{abstract}

\section*{Introduction}

The search for New Physics beyond the Standard Model is one of the main goals of high-energy physics.
A fairly common strategy is to search for a localized deviation in an invariant mass spectrum that could correspond to a new heavy particle.
This kind of search usually depends on accurate simulations of the Standard Model processes and also on several signal hypotheses.
However, simulating data from experiments such as ATLAS \cite{ATLAS} is computationally intensive and is limited by modelling uncertainties.
Also, assuming a signal model without knowing what lies beyond the Standard Model can be a source of bias that reduces the generalizability of an analysis.

To overcome these limitations, much effort has been put into defining generic search strategies that do not rely on specific theoretical models of New Physics.
One possible solution is to use algorithms that don't need a specific signal model to train on, but still detect events that differ from the Standard Model predictions.
Such unsupervised anomaly detection algorithms~\cite{anom_det} can potentially identify anomalous events by evaluating an anomaly score, so that in the search for New Physics processes, 
signal events can be seen as an anomaly with respect to the Standard Model.

A well-known class of anomaly detection algorithms using unsupervised machine learning is the auto-encoder (AE) and its derivatives \cite{AE1,AE2}.
Such models can be trained directly on data with the only assumption that signal events are very rare.
In the following sections, we present a GAN-AE algorithm inspired by AEs and generative models that allows for both anomaly detection and data-driven background modeling. 
This model is tested on the LHC Olympics 2020 challenge dataset \cite{LHCO} as a benchmark.
For this search a complete strategy including the model independent BumpHunter algorithm \cite{BH2011} has been defined.
The code used to build and train the GAN-AE algorithm on this dataset is accessible online\footnote{\href{https://github.com/lovaslin/GAN-AE}{https://github.com/lovaslin/GAN-AE}}.

\section{The GAN-AE algorithm}
\label{sec:GAE}

\begin{figure}[ht]
    \centering
    \includegraphics[width=0.8\linewidth]{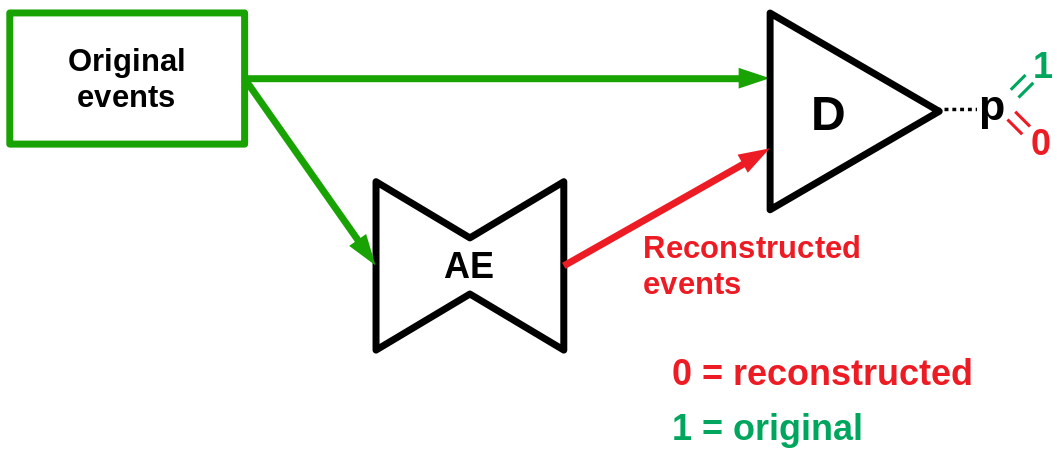}
    \caption{Schematic of the global layout of the GAN-AE architecture.
    The auto-encoder network (AE) is trained to produce reconstructed events that closely resemble the original events.
    The discriminator network (D) is trained to discriminate between reconstructed and original events with labels 0 and 1, respectively.}
    \label{fig:GAE}
\end{figure}

The GAN-AE algorithm proposes to combine a vanilla auto-encoder together with a discriminator network in an adversarial manner similar to that of a Generative Adversarial Network (GAN) \cite{GAN}.
Other algorithms propose similar models, such as Outliers Exposure \cite{OE_AE} and Self-Adversarial AE \cite{adVAE}.
In these works, the goal is either to constrain the latent space of an AE or to improve the sensitivity to anomalies in a semi-supervised setting.
With the GAN-AE algorithm, the objective is to construct an alternative measure of reconstruction error using a multilayer perceptron network trained to distinguish reconstructed and original events.
Figure~\ref{fig:GAE} shows a synoptic view of the GAN-AE architecture.

Traditionally, auto-encoders are trained using a possibly regularized measure of the (Euclidean) distance between their input and output.
A well known metric for this task is the Mean Square Error (MSE).
In this work, we propose an alternative metric based on a supervised discriminator network trained to classify reconstructed events (labeled 0) and original events (labeled 1).
This binary classifier (bc) model is trained with the usual binary cross-entropy loss function:
\begin{equation}
    \text{bc}\left(y^{(d)},y^{(l)}\right)=- \left[y^{(l)} \log \left(y^{(d)}\right)
    +\left(1-y^{(l)}\right) \log \left(1-y^{(d)}\right) \right]\, ,
    \label{eq:bc}
\end{equation}
where $y^{(d)}$ is the output of the discriminator and $y^{(l)}$ the associated label.

In order to train this two-party GAN-AE network, we define a training procedure divided into two main phases.
The first step is to train the discriminator network parameters $\boldsymbol{{\theta}_D}$ with a mixture of original data and events reconstructed by the AE.
Parameters $\boldsymbol{{\theta}_D}$ are then updated for a few epochs while keeping the parameters $\boldsymbol{{\theta}_{AE}}$ of the AE fixed.

The second step is to train the auto-encoder parameters $\boldsymbol{{\theta}_{AE}}$ using the discriminator output as constraint.
This training is done with a special loss function that combines both the usual distance metric and the information coming from the discriminator.
The distance metric used is a modified Euclidean distance defined as:
\begin{equation}
    \text{D} \left(y^{(o)},y^{(r)}\right)=
    \sqrt{\frac{1}{N} \sum_{i=1}^{N}\left(y_i^{(o)}-y_i^{(r)} \right)^2 }\, ,
    \label{eq:dist}
\end{equation}
with $y^{(o)}$ the input vector (original event), $y^{(r)}$ the output vector (reconstructed event) and $N$ the dimension of both vectors. The constraint of the discriminator is introduced by modifying the binary cross-entropy loss function defined in equation \ref{eq:bc}.
In fact, while the goal of the discriminator is to correctly identify reconstructed events associated with the label `0', the goal of the AE is, on the contrary, to confuse the discriminator network.
Thus, the AE must be trained so that the output of the discriminator comes closer to the label `1' corresponding to (real) original events.
This can be achieved by computing the binary cross-entropy loss of the discriminator using reconstructed events associated with the label of the original events as the target.
The two metrics are then combined to define the loss for a given event $k$ as follows:
\begin{equation}
    \text{L}_k \left(y^{(o)},y^{(r)},y^{(d)}\right)=
    \text{bc} \left(y^{(d)}, y^{(l)}=1 \right)
    +\varepsilon\text{D} \left(y^{(o)},y^{(r)}\right) \, ,
    \label{eq:loss_AE}
\end{equation}
with $\varepsilon$ a hyperparameter that balances the relative importance of the two terms.
This loss is used to update $\boldsymbol{{\theta}_{AE}}$ for a few epochs.

The AE has an architecture composed of 5 layers: the encoding part with the input layer, a hidden layer and the latence space, and a decoding part that is exactly symmetrical to the encoder part.
The activation function used for the hidden layers is the LeakyReLU function, while the latent space and output are linear.
As an additional constraint, we used the tied weight trick discussed in \cite{TiedW} to impose that the weight tensors of the decoder are the transposed ones of those of the encoder:
\begin{equation}
    W^{(n-k)} = \left(W^{(k)}\right)^{\text{T}} \, ,
    \label{eq:teidW}
\end{equation}
where $W^{(k)}$ is the weight tensor between layers $k$ and $k+1$ of the encoder and  $W^{(n-k)}$ is the weight tensor between layers $n-k$ and $n-k-1$ of the decoder.
Dropout is applied to each hidden layers.

The structure of the discriminator network is defined as a fully connected multilayer perceptron with 4 hidden layers using LeakyReLU activation.
The output is one-dimensional with a sigmoid activation function compatible with the binary cross-entropy loss function.
Dropout is applied to the hidden layers of the discriminator.

The main hyperparameters of the GAN-AE algorithm are presented in Section~\ref{sec:opti}. In this architecture, the discriminator is used to enhance the training of the auto-encoder. 
However, in the application step, only the trained AE is actually used.
The anomaly score is defined as the modified Euclidean distance (equation \ref{eq:dist}).
Thus, the most anomalous events, here assimilated to the most signal-like events, can be identified as those with the highest anomaly score.
The selected anomalous events can then be compared to a reference to test for the presence of an anomaly.
The next section describes how to obtain this reference.

\section{Background modelling and mass sculpting mitigation}
\label{sec:bkg_mod}

In order to integrate the GAN-AE algorithm into a complete and fully data-driven search strategy, we propose a method to extract a viable background model directly from the data.
This method is based on the hypothesis that the signal that we might expect to find in the data is a rare process, such that the data is dominated by the background.
In this case, when performing a bump hunt in a relevant spectrum, such as an invariant mass, one would expect the signal to be invisible unless proper selections are made.
Thus, the invariant mass spectrum prior to any selection can serve as the reference background distribution.

\begin{figure}[ht]
    \centering
    \includegraphics[width=0.75\linewidth]{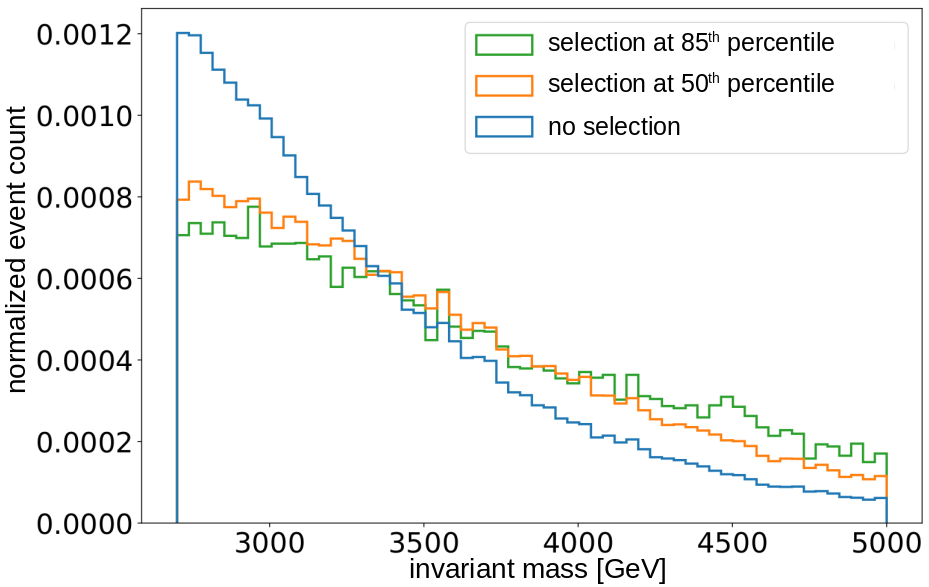}
    \caption{Normalized histograms of the invariant mass.
    The blue histogram shows the spectrum before applying any selection to the anomaly score.
    The orange and green histograms show the spectra after selection at the 50th and 85th percentiles of the anomaly score distribution, respectively.
    The data used to obtain this figure is described in Section \ref{sec:LHCO}.
    }
    \label{fig:mass_sculpt}
\end{figure}

However, in order to use this distribution as a reference background, we must ensure that its shape is not affected by the selection based on the anomaly score described in the previous section.
Even if the GAN-AE model is trained without using the invariant mass as an input variable, this condition is generally not met, as illustrated in Figure \ref{fig:mass_sculpt}.

To get rid of the mass sculpting induced by the selection process, we propose two mitigation techniques that can be combined.
First, an event weight is applied in order to uniformize the invariant mass distribution. 
Otherwise, events with low invariant mass will be over-represented in the data compared to others, leading to a bias in the reconstruction error.
Then, to reduce the mass sculpting, the Distance Correlation regularization (DisCo)~\cite{DisCo1, DisCo2} is added to the loss of the auto-encoder.
As it requires the use of independent and identically distributed samples of the distribution to decorrelate, this term is defined for a batch of events.

By combining the DisCo regularization term 
and the event weighting, we can define the modified loss function of the auto-encoder:
\begin{equation}
    \text{L}_k \left(y^{(o)},y^{(r)},y^{(d)}\right)=
    \sum_{i=1}^{N_b} w_i \left[ \text{bc} \left(y_i^{(d)}, y_i^{(l)}=1 \right)
    +\varepsilon\text{D} \left(y_i^{(o)},y_i^{(r)}\right) \right] \\
    +\alpha\text{DisCo}\left(\text{D}_{\Sigma},y^{(m)} \right)
    \label{eq:loss_full}
\end{equation}
with $w_i$ the weight associated to event $i$, $N_b$ the number of events in a batch, $\alpha$ a new hyperparameter of the loss, $y^{(m)}$ the vector of invariant mass value associated to a batch and $\text{D}_{\Sigma}$ the vector of anomaly score values associated to a batch.
Note that the event weights should not be applied when computing the DisCo regularization.
Since the goal of this term is to decorrelate the invariant mass and anomaly score distributions, it is important to keep both distributions unchanged.

With this new loss function, we can ensure that the invariant mass distribution prior to the selection on the anomaly score is a valid reference model for the background.
Now we need to compare this reference with the distribution of selected events in order to look for a localized deviation.
For this purpose we use the pyBumpHunter package \cite{pyBH2}\footnote{\href{https://github.com/scikit-hep/pyBumpHunter/tree/v0.4.0}{GitHub - scikit-hep/pyBumpHunter at v0.4.0}} which provides an improved version of the BumpHunter algorithm \cite{BH2011} implemented in Python.
This tool has the advantage of locating any deviation in a model-independent way, evaluating both local and global significance by removing the Look Elsewhere Effect \cite{LEE}.
Now we have all the tools needed to build a complete and model-independent strategy for resonant New Physics searches.
The next section shows an example of application using a benchmark dataset.

\section{Application to LHC Olympics 2020 data}
\label{sec:LHCO}

\begin{figure}[ht]
    \centering
    \includegraphics[width=0.95\linewidth]{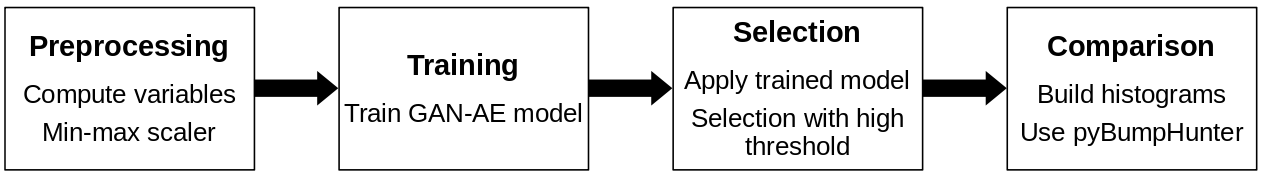}
    \caption{Diagram representing the analysis flow applied for the LHC Olympics 2020 challenge.}
    \label{fig:strat}
\end{figure}

In order to test and evaluate the performance of the techniques developed in the previous section, we use the public dataset proposed for the LHC Olympics 2020 challenge~\cite{LHCO}.
This dataset provides a good case study for testing and comparing anomaly detection algorithms in the context of model-independent New Physics searches.
The strategy that we will use for this challenge is illustrated in figure~\ref{fig:strat}.

The challenge proposes a so-called RnD dataset \cite{LHCO_data} to assist the development of anomaly detection algorithms. This dataset is composed of a background sample containing QCD multijet events and a benchmark New Physics signal model.
The signal events consist of a Z' boson with a mass of 3.5 TeV (inspired by \cite{Zprime}) decaying into two heavy resonances X and Y with masses of 500 GeV and 100 GeV, respectively.
Two types of signal signatures are considered, one where both X and Y decay to two quarks and form boosted jets with 2-pronged substructure, 
and another where both X and Y decay to three quarks, resulting in boosted jets with 3-pronged substructure.
A total of 1M events were generated for the background model, along with 100k events for each signal hypothesis. 
The events are generated using \textsc{Pythia}8  \cite{Pythia} and \textsc{Delphes}~3.4.1 \cite{Delphes} with no pile-up or multiple parton interaction included, and with a detector architecture similar to the ATLAS experiment.
Events are selected using a large radius (R=1) jet trigger with a $p_T$ threshold of 1.2 TeV. 

The anomaly detection algorithms are tested on three different Black Box datasets~\cite{LHCO_data2} containing unknown event samples.
The only information given to the challenge participants is that the events contain at least two jets with a different background modelling than the RnD data.
The goal is then to determine if there is a hidden signal in the Black Boxes and at what mass.

For each event, a list of up to 700 hadrons 4-vectors is provided.
Jets are reconstructed using the anti-$k_t$ algorithm implemented in the \textsc{FastJet}~3.3.3 library \cite{FastJet} with a large jet radius $R=1$.
A second clustering is performed within the large jets with a smaller radius $r=0.2$ in order to characterize their substructure.
The list of the variables computed in this preprocessing procedure is presented in Table~\ref{tab:LHCO_var}.
For a clustering in two large jets, we have a total of 45 variables.
The code used to preprocess the data is publicly available\footnote{\href{https://gitlab.cern.ch/idinu/clustering-lhco}{https://gitlab.cern.ch/idinu/clustering-lhco}}.

\begin{table}[ht]
    \centering
    \begin{tabular}{|c|c|c|}
        \hline
        4-vectors & $p_T$, $\eta$, $\phi$, $E$ \\
        \hline
        Jet mass and constituents & $m_{jet}$, $n_c$ \\
        \hline
        Number of subjets~\cite{FastJet}  &  $N_{incl}$, $N_{excl}$ \\
        \hline
        N-subjetiness \cite{Nsubj}  &  $\tau_1$, $\tau_2$, $\tau_3$, $\tau_{21}$, $\tau_{31}$ \\
        \hline
        Energy Rings  &  $E_{ring,1}$, $E_{ring,2}$, ... , $E_{ring,10}$ \\
        \hline\hline
        Di-jet invariant mass &  $mjj$ \\
        \hline
    \end{tabular}
    \caption{Table summarizing all variables, computed in the preprocessing of the LHC Olympics 2020 data,
    for each large jet, except for the last variable, which is defined for pairs of jets.
    }
    \label{tab:LHCO_var}
\end{table}

\subsection{Hyperparameter optimization}
\label{sec:opti}

The hyperparameters are optimized according to three Figures of Merits (FoM).
The first is the reconstruction error of the AE network, which also serves as an anomaly score.
The performance being maximal when the reconstruction error is minimal.
Another important criterion, discussed above, is the amount of mass sculpting when applying a cut to the anomaly score distribution.
Finally, a third aspect to consider is the stability of the method.
This is necessary to ensure the resilience of the algorithm to random fluctuations in the initialization of the model parameters prior to training.
It can be evaluated by training a model several times with the same hyperparameters but with different initialization seeds.
All these criteria should be taken into account in order to optimize the GAN-AE in a model-independent way.

In the context of the present work, we chose an initial set of hyperparameters based on our previous expertise with Auto-Encoders and GANs.
These hyperparameters, used as starting point for the optimization procedure, are reported in Table \ref{tab:hyper_ini}.
\begin{table}[ht]
    \centering
    \begin{tabular}{|c|c||c|c|}
        \hline
        Hyperparameter & Value  &  Hyperparameter  & Value \\
        \hline
    Latent space dimension & 14  &  Number of training cycles  &  100  \\
        \hline
        AE hidden dimension & 84  &  AE epochs per cycle  &  5  \\
        \hline
        D hidden dimensions & \{150, 100, 50\}  &  D  epochs per cycle  &  7  \\
        \hline
        $\varepsilon$ (reconstruction term) & 5.0  &  Pre-training of AE  &  True  \\
        \hline
        $\alpha$ (DisCo term) & 50.0  &  Dropout rate &  20\%  \\
        \hline
    \end{tabular}
    \caption{Initial set of hyperparameter used at the begining of the optimization procedure.
    A pre-training of the Auto-Encoder is performed without adversary using only the reconstruction error before the main training loop.}
    \label{tab:hyper_ini}
\end{table}

The different FoM of the model can be evaluated with these initial values and then at neighboring points in the hyperparameter space.
The position of the points is then updated according to the results.
This technique is similar to that used to update the centroids defined in the well-known k-means algorithm.
However, the selection of the best hyperparameter set is difficult to automate. 
Therefore, based on this procedure, the hyperparameters were optimized empirically.

Compared to Table \ref{tab:hyper_ini}, the architecture of the discriminant was changed to four hidden layers with dimensions \{300, 200, 100, 50\}.
Also, the hyperparameters of the loss function were changed to 6.0 for $\varepsilon$ and 65.0 for $\alpha$.
Increasing the DisCo regularization term was found to be important to reduce the mass sculpting.

\subsection{Results on RnD data}

In order to evaluate the performance of the GAN-AE algorithm and validate the background modeling procedure, we use the RnD dataset.
The results are presented for a clustering in two large jets.
The GAN-AE model is trained on 100k background events and tested on a mixture of background and signal.
All variables listed in Table~\ref{tab:LHCO_var} are used in the training except for the di-jet invariant mass and the azimuthal angle $\phi$ of the jets, for a total of 42 input variables.

\begin{figure}[ht]
    \centering
    \begin{subfigure}[t]{.46\textwidth}
    \centering
    \includegraphics[width=\linewidth]{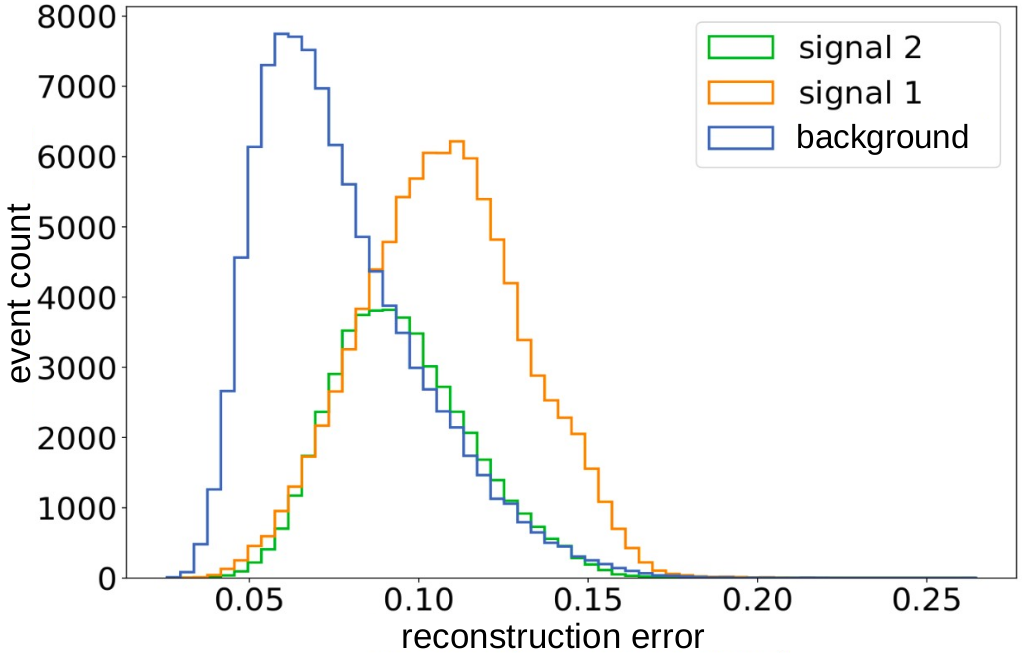}
    \caption{}
    \label{fig:RnD_dist}
    \end{subfigure}
    \,
    \begin{subfigure}[t]{.46\textwidth}
    \centering
    \includegraphics[width=\linewidth]{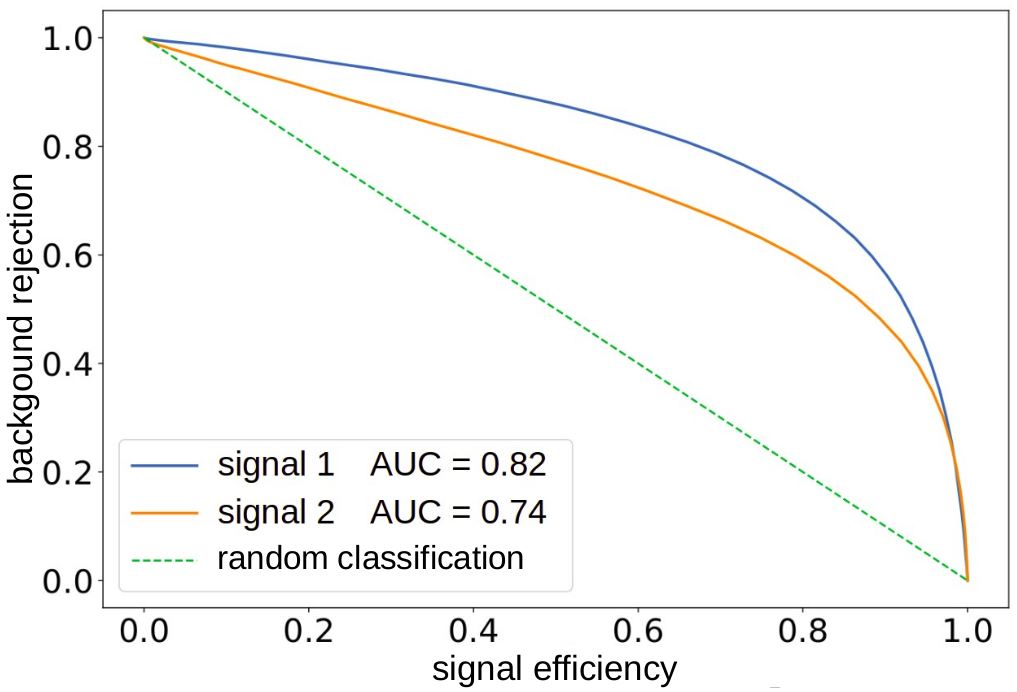}
    \caption{}
    \label{fig:RnD_ROC}
    \end{subfigure}
    \caption{Results obtained with the RnD data of the LHC Olympics 2020 challenge showing the separation of background and signal: (a) anomaly scores for background and signal events; (b) ROC curves obtained from the test set of the RnD data. The labels signal~1 (orange) and signal~2 (green) correspond to 2-prong and 3-prong jet substructure, respectively.}
    \label{fig:RnD1}
\end{figure}

The anomaly scores obtained for the background and both signal test samples are shown in Figure~\ref{fig:RnD_dist}.
The corresponding ROC curves are shown in Figure~\ref{fig:RnD_ROC}.
The Area Under the Curve (AUC) obtained on the test set is 0.82 for the first RnD signal (2-prong) and 0.74 for the second (3-prong)
This result confirms that the Auto-Encoder trained using the GAN-AE algorithm is able to distinguish the signal from the background.

\begin{figure}[ht!]
    \centering
    \begin{subfigure}[t]{.46\textwidth}
    \centering
    \includegraphics[width=\linewidth]{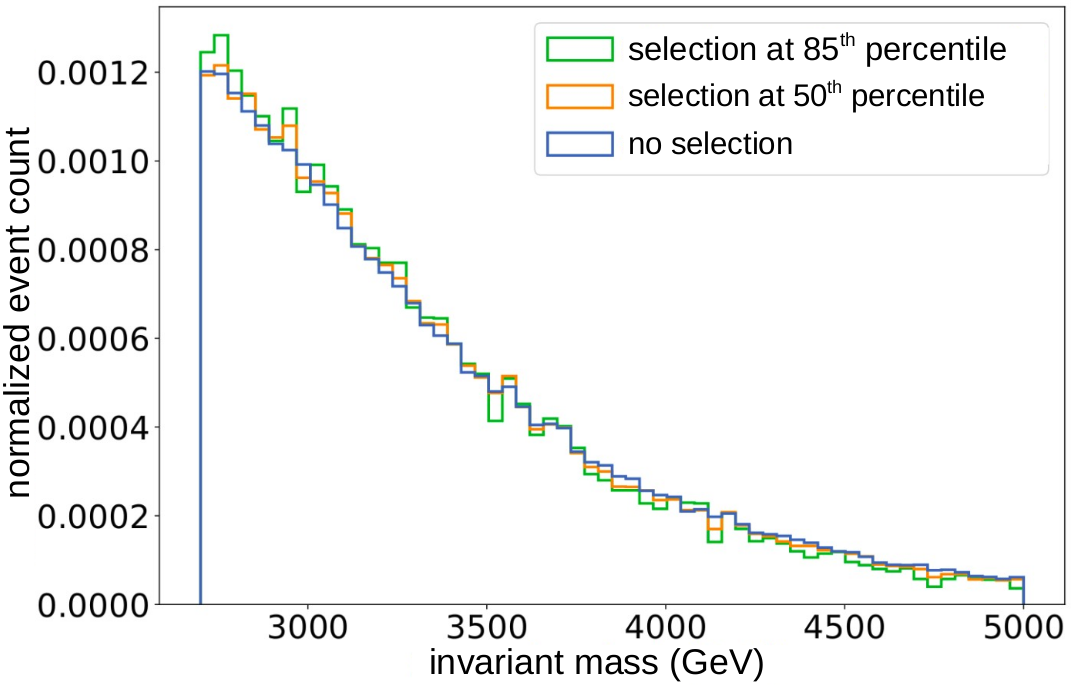}
    \caption{}
    \label{fig:RnD_mass}
    \end{subfigure}
    \,
    \begin{subfigure}[t]{.46\textwidth}
    \centering
    \includegraphics[width=\linewidth]{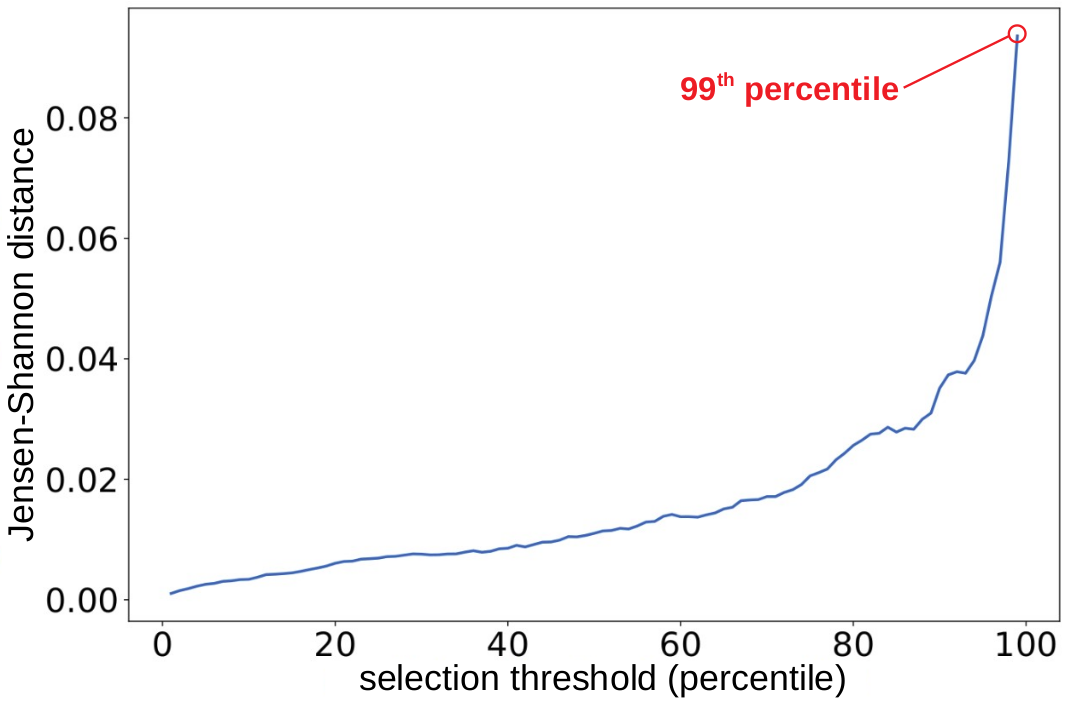}
    \caption{}
    \label{fig:RnD_JSD}
    \end{subfigure}
    \caption{Results obtained with the RnD data of the LHC Olympics 2020 challenge showing the capacity to mitigate the mass sculpting:
    (a) di-jet invariant mass of background events before (blue) and after selection at the $50^{th}$ (orange) and $85^{th}$ (green) percentile of the anomaly score distribution, (b) Jensen-Shannon divergence between the invariant mass distribution before and after selection for different thresholds.
    }
    \label{fig:RnD2}
\end{figure}

To check the ability to remove mass sculpting, the modeling of the reference background distribution in the test set is evaluated after applying a selection on the anomaly score.
Figure~\ref{fig:RnD_mass} shows the normalized distribution of the di-jet invariant mass, before and after selection at different thresholds.
To quantitatively assess the deformation of the invariant mass spectra induced by the selection, we use Jensen-Shannon divergence as a metric \cite{JSD}.
By continuously varying the selection threshold, we can evaluate this metric to produce the curve shown in Figure~\ref{fig:RnD_JSD}.
Compared to the results shown on Figure~\ref{fig:mass_sculpt}, the invariant mass distribution is no longer modified when applying a selection based on the anomaly score.
The fact that Jensen-Shannon divergence stays below 0.1 up to a $99^{th}$ percentile threshold indicates that the invariant mass distribution of the background before selection remains compatible with that after selection.

\begin{figure}[ht!]
    \centering
    \includegraphics[width=0.75\linewidth]{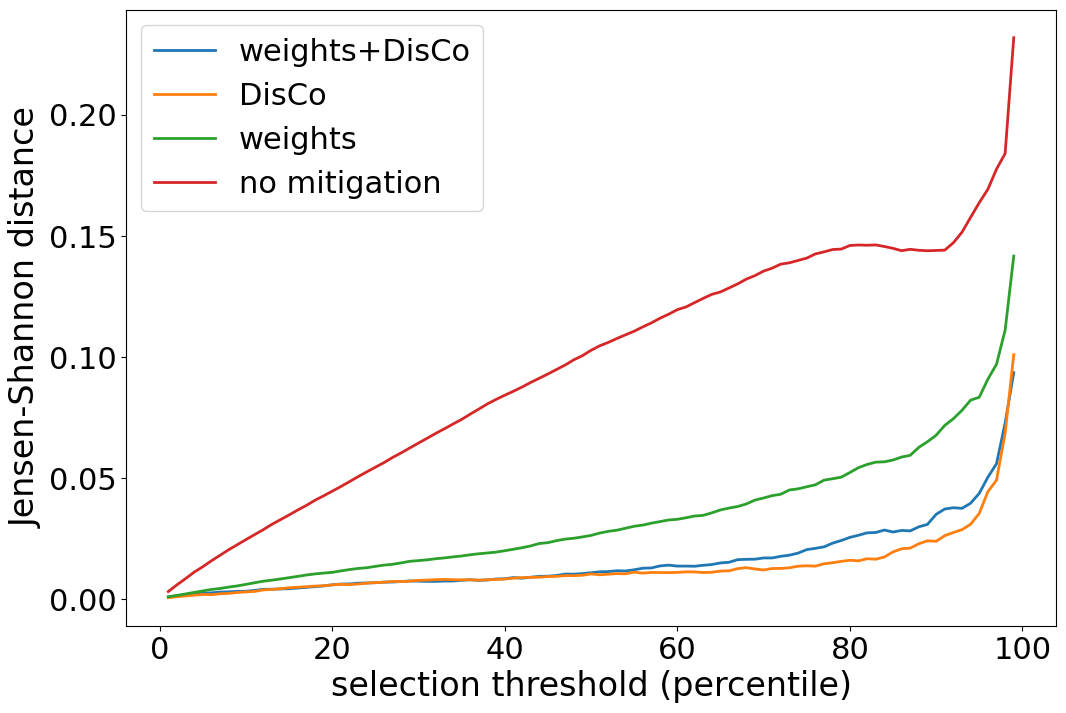}
    \caption{Jensen-Shannon distance obtained using the two mass sculpting mitigation techniques (blue) and without using them (red). The orange and green lines correspond, respectively, to the Jensen-Shannon distance obtained using only DisCo and only event weights.}
    \label{fig:RnD_comp}
\end{figure}

By comparison, a GAN-AE model trained without the mass sculpting mitigation techniques results in the Jensen-Shannon divergence curve shown in Figure~\ref{fig:RnD_comp}.
This metric increases rapidly with the selection threshold reaching more than twice the distance obtained with the mitigation techniques.
This strong constraint on the mass sculpting can be realized simultaneously with the good ability to separate signal and background shown in Figure~\ref{fig:RnD1}.
This achievement is a good improvement over classically trained Auto-Encoders for which applying such constraints generally deteriorates the quality of the anomaly detection.

Figure~\ref{fig:RnD_comp} also shows the results obtained using only one mitigation technique.
Using the DisCo regularization results in a Jensen-Shannon distance that is similar, within uncertainties, to that obtained with both mitigation techniques.
However, when using only the DisCo regularization, we observed a localized deviation at high selection thresholds, which appeared at a low invariant mass.
This deviation can be identified as a spurious signal when performing a model-independent bump hunt.
Adding event weights is the only way we found to reduce this spurious signal without affecting the anomaly detection performance.

\begin{figure}[ht!]
    \centering
    \includegraphics[width=0.65\linewidth]{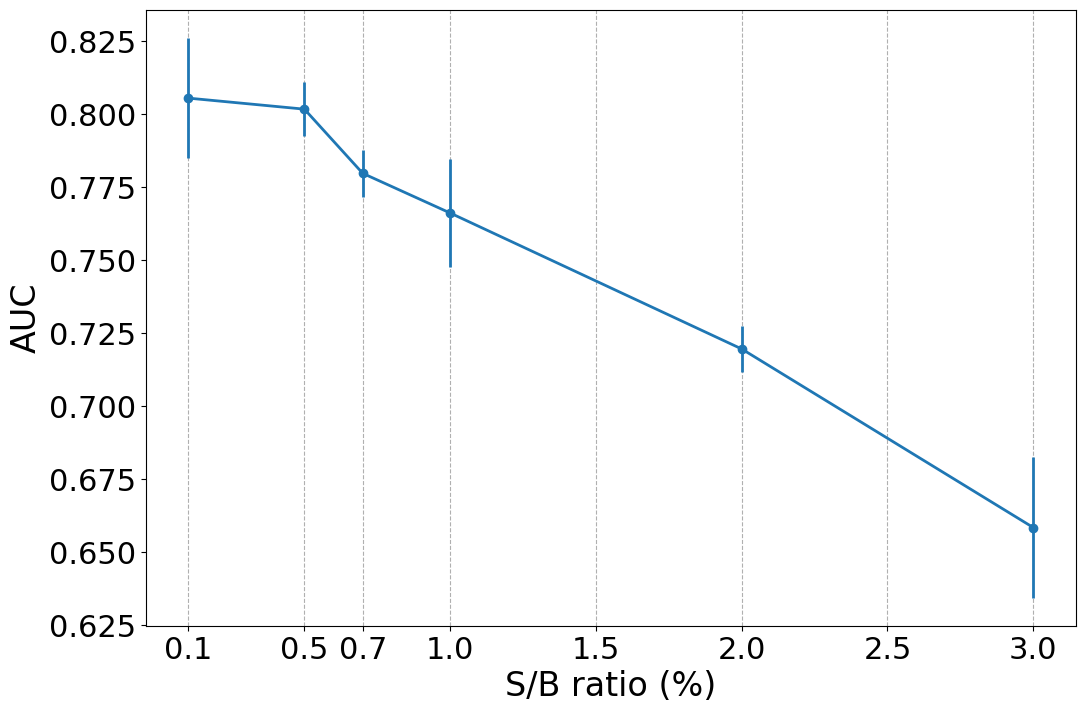}
    \caption{Evolution of the average AUC as function of the signal fraction in the data. Error bars are obtained by training multiple models with the same signal fraction.}
    \label{fig:RnD_inject}
\end{figure}

Finally, one last check concerns the effect of signal contamination in the training set. 
The main hypothesis behind our strategy is that the signal should be very small compared to the background~\footnote{
This hypothesis seems valid if we consider the limits obtained at the LHC for di-jet resonance searches, for example in~\cite{dijet_ATLAS} or~\cite{dijet_CMS}. 
}.
To quantify up to which signal fraction this hypothesis can hold, we performed signal injection tests using the first RnD signal (2-pronged substructure).
In this process, the amount of signal in the training set is gradually increased, and a new model is trained at each step.
We have found that the AUC obtained using the same hyperparameters as presented above is stable up to a signal over background (S/B) ratio of 0.5\%, and remains satisfactory up to 1\% with an AUC above 0.75.
Beyond this value, the AUC deteriorates rapidly with increasing signal fraction. Figure \ref{fig:RnD_inject} shows the evolution of the AUC for signal fraction values up to 3\%.

\subsection{Results on Black Box datasets}

After validating the GAN-AE algorithm and the mass sculpting mitigation procedure, we can apply the complete strategy chain to the Black-Box datasets.
For each Black-Box, a GAN-AE model is trained on 100k events using the set of hyperparameters presented in Section~\ref{sec:opti}.
The trained model is applied to each dataset in order to evaluate the anomaly score distribution.
A selection is applied on the $99^{th}$ percentile of this distribution.
Then, the invariant mass distribution of the di-jets in this subsample is compared to the invariant mass distribution of the di-jets in the pre-selection data, which serves as a reference background.
The reference histogram is normalized to the selected data using a side-band normalization procedure.

\begin{figure}[ht]
    \centering
    \begin{subfigure}[t]{.46\textwidth}
    \centering
    \includegraphics[width=\linewidth]{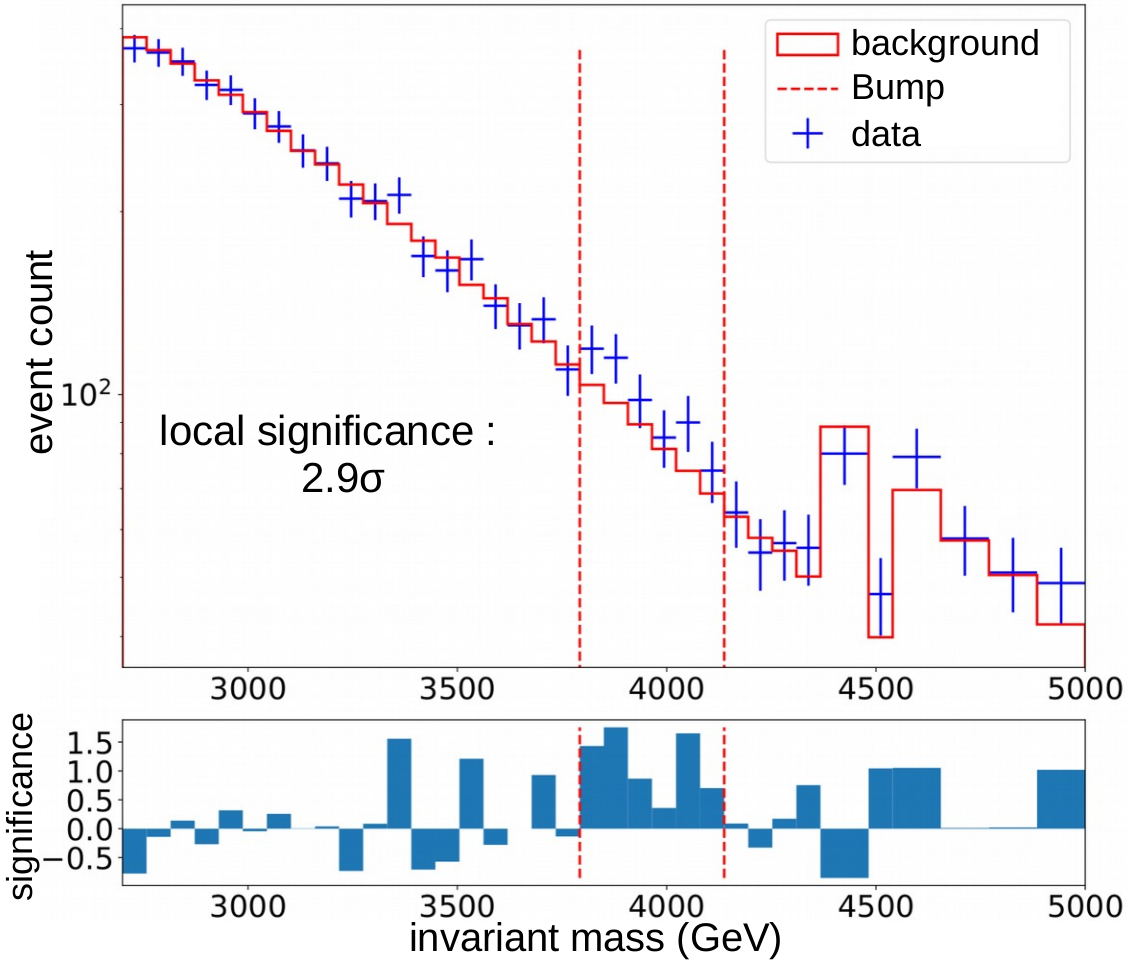}
    \caption{In the top panel, histograms of di-jet invariant mass after selection (blue) and reference background (solid red).
    In the bottom panel, local significance per bin of the invariant mass histograms.
    The vertical dashed lines represent the interval selected by the BumpHunter algorithm.}
    \label{fig:BB1_mass}
    \end{subfigure}
    \,
    \begin{subfigure}[t]{.46\textwidth}
    \centering
    \includegraphics[width=\linewidth]{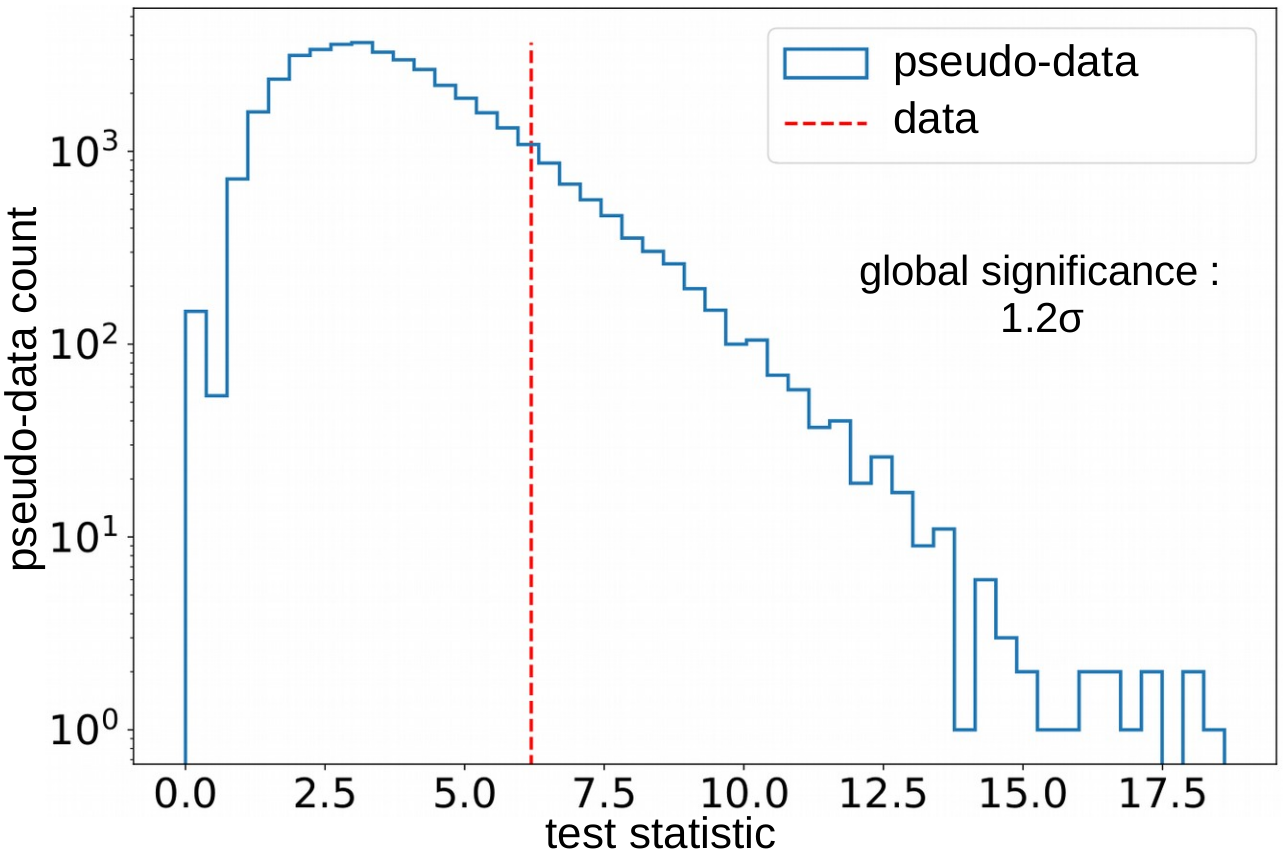}
    \caption{Distribution of test statistic obtained for background-only pseudo-data (blue histogram), together with the value obtained for observed data (dashed line).}
    \label{fig:BB1_global}
    \end{subfigure}
    \caption{Results obtained with the data of Black-Box~1 of the LHC Olympics 2020 challenge after applying the complete analysis chain.}
    \label{fig:BB1_pyBH}
\end{figure}

Results obtained with pyBumpHunter for Black-Box~1 are presented in Figure~\ref{fig:BB1_pyBH}.
The BumpHunter algorithm finds a deviation in data, with respect to the data-driven reference background, around 3.97 TeV with a local significance of almost~$3\sigma$ (Figure~\ref{fig:BB1_mass}).
No other significant excess, or deficit, is observed outside the selected interval.
Figure~\ref{fig:BB1_global} shows the background-only test statistics from which a global significance of $1.2\sigma$ is derived.
The low overall significance is partly explained by the fact that the bump hunt search is performed without assuming a prior signal and with a floating background normalization.

\begin{figure}[ht]
    \centering
    \includegraphics[width=0.75\linewidth]{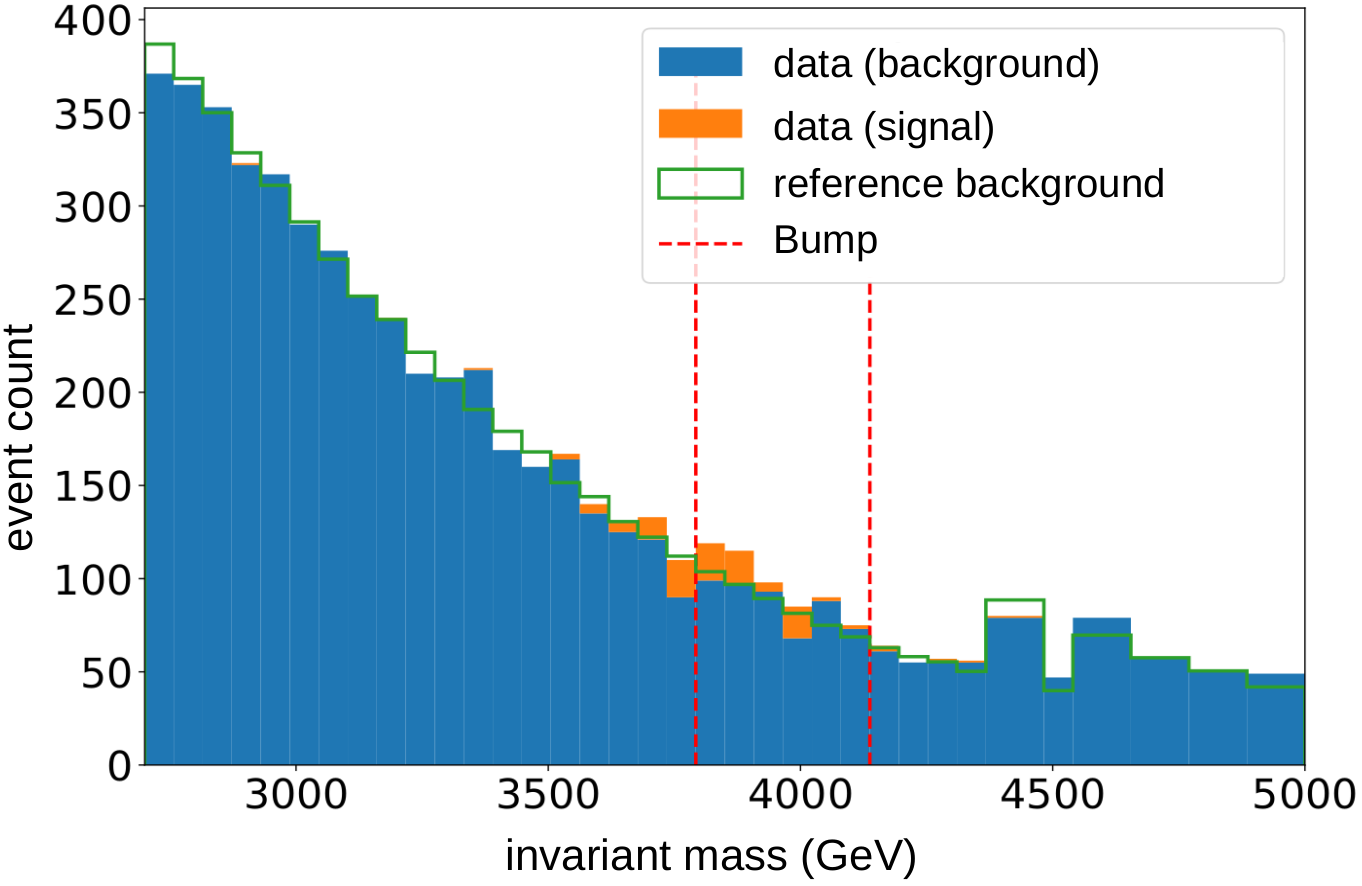}
    \caption{Histograms showing the true background (blue) and signal (orange) distributions for Black-Box~1, after selection on the $99^{th}$ percentile of the anomaly score distribution.
    The reference background histogram used for BumpHunter is shown in green and the selected interval is represented by the vertical dashed lines.}
    \label{fig:BB1_truth}
\end{figure}

After the end of the challenge, the content of each Black-Box was revealed by the organisers.
Figure~\ref{fig:BB1_truth} shows the histograms of di-jet invariant mass in Black-Box~1, along with the true labels
corresponding to background and signal events. 
The region of the spectrum identified by the BumpHunter algorithm is indeed the location of the true signal.
The signal generated for this dataset corresponds to a 3.8~TeV Z' boson decaying to two heavy resonances with a similar 2-prong substructure jet signature as in the RnD data.

The initial signal over S/B ratio is 0.08\%, which is within the application range of our method.
After applying the full strategy chain to this dataset, we obtain an improvement in the S/B ratio of a factor of~20.
The signal efficiency after selection at the $99^{th}$ percentile of the anomaly score distribution is over 15\% for a background rejection of almost 99\%.
We also note that the data-driven reference background fits quite well the true background distribution after selection.
The deviation identified by BumpHunter corresponds to the true signal with a small bias on the mass of the Z' (less than 200 GeV).

The same methodology has been applied to the two other Black-Boxes and results are summarized below.
Black-Box~2 did not contain any signal, as this data set was actually provided for the purpose of testing the identification of spurious signals.
Our algorithm successfully modeled the shape of the background and found no significant deviations. 
The third black box contained a complex signal signature, as the generated resonance could decay into either two or three jets, with a branching ratio of one third and two thirds, respectively. In the case of Black-Box~3 and with the 2-jet clustering, the GAN-AE algorithm was unable to distinguish between signal and background events. However, the process of modeling the background shape from the data still worked.

Finally, we compared the performance of the GAN-AE algorithm with a classical supervised algorithm using a BDT model implemented in scikit-learn~\footnote{\href{https://scikit-learn.org/stable/index.html}{https://scikit-learn.org/stable/index.html}}.
The model was trained on the RnD background and 2-pronged signal and applied to Black-Box~1 sample.
The AUC obtained on the RnD dataset is 0.99, but it drops to 0.89 when applied to the Black-Box dataset.
We also evaluated an approximate local significance~\footnote{We use the approximation $\sigma=S/\sqrt{S+B}$ with S and B the number of signal and background events respectively.} of 3.5$\sigma$ considering a selection threshold at the 99th percentile of the BDT output distribution.
In comparison, the AUC obtained for a GAN-AE trained directly on the Black-Box dataset is 0.85 and the local significance is 2.9$\sigma$.
In addition, since the BDT does not use the DisCo regularization or the event weights, there is no control over the mass sculpting.
Thus, there is no easy way to extract a suitable background model to perform a bump hunt, other than using simulated Monte-Carlo samples.

\section*{Conclusion}

The development of alternative search strategies for New Physics beyond the Standard Model has gained much importance in recent years.
Events such as the LHC Olympics challenge proposed in 2020 are part of this effort.
In this context, we propose a model-independent analysis strategy based on unsupervised machine learning and data-driven background modeling.

The GAN-AE algorithm offers an interesting alternative to the classical training of auto-encoders by defining a new measure of reconstruction error given by an adversary network.
This algorithm offers good performance and stability, even when using strong constraints to reduce the mass sculpting such as the DisCo regularization term.
Thanks to this constraint, we can derive a reference background model directly from the data, with the only assumption that the signal is rare enough.
The background model can then be used as a reference for the BumpHunter algorithm, which allows the evaluation of both local and global significance.

The strategy was tested using the LHC Olympics 2020 challenge datasets.
The results on the RnD dataset as well as on the first black box are promising, allowing us to correctly identify the hidden signal with a local significance of $2.9\sigma$.
This result is comparable to those obtained by other participants.
Our strategy is also the only one to propose a built-in evaluation of the global significance, showing its completeness.
A possible way to improve the method would be to include the GAN-AE algorithm in a weakly supervised setting, such as the Tag N'Train (TNT) algorithm \cite{TNT}, which obtained one of the best results in the LHC Olympics 2020 challenge.

\section*{Acknowledgement}

The authors would like to thank Gregor Kasieczka, Ben Nachman, and David Shih, the organizers
of the LHC Olympics 2020 Anomaly Detection Challenge, for providing the datasets used in this
study and for the opportunity to develop and test the GAN-AE architecture.\\

\noindent Louis Vaslin acknowledges the support received by the French government IDEX-ISITE initiative 16-IDEX-0001 (CAP 20-25).

\bibliography{ref}
\bibliographystyle{unsrt}

\end{document}